\documentclass[a4paper]{article}
\usepackage[utf8]{inputenc}
\usepackage{authblk}
\usepackage{setspace}
\usepackage[margin=1.25in]{geometry}
\usepackage{graphicx}
\graphicspath{ {./figures/} }
\usepackage{subcaption}
\usepackage{amsmath}
\usepackage{xcolor}
\usepackage[english]{babel}
\usepackage[utf8]{inputenc}
\usepackage{csquotes}

\usepackage[style=nejm, 
citestyle=numeric-comp,
sorting=none]{biblatex}
\bibliography{main.bib} 
\title{Statistical analysis of IUVB values in Mexico City from 2000 to 2022}

\author[1,2*]{C. H.~Zepeda~Fern\'andez}
\author[1]{F. A. S\'anchez-Ar\'evalo}
\author[2]{E. Moreno~Barbosa}

\affil[1]{Consejo Nacional de Humanidades Ciencia y Tecnolog\'ia}
\affil[2]{Facultad~de~Ciencias~F\'isico Matemáticas,~Benem\'erita~Universidad~Aut\'onoma~de~Puebla}
\affil[*]{Address correspondence to: hzepeda@fcfm.buap.mx}

\date{}

\begin{document}

\maketitle

\begin{abstract}
The solar radiation are electromagnetic waves, composed of infrared, visible spectrum and ultraviolet. The infrared component is the cause of thermal energy, the visible spectrum allows to see and the ultraviolet component is the most energetic part and dangerous for the human body (skin and eyes). The ultraviolet rays are divided in a wavelength  range, in three parts; called UVA (100-280~nm), UVB (280-315~nm) and UVC (315-400~nm). The UVC are the most energetic (followed by the UVB rays ), however are stopped in the ozone layer. In this work it shown a statistical analysis of the UVB index measured by five station in Mexico City, from  the years 2000 to 2022. Through a Gaussian fit distribution,  it was found that the range when the IUVB value exceeds the value of seven, which is from 11:00 to 16:00, having an  average mean value of 13:09~hrs.~$\pm$~4~min. i.e. the time at which the UVB index reaches its maximum value. This occurs in the months from February to October.   More than 80\% of  radiation is for UVB values less than 7 and less than the remaining 20\% is for IUVB values  greater than 7. Finally, it was possible to observe that the  mean value per year, reaches its maximum when the solar cycles occurs, which was in the years 2003 and 2013.

\end{abstract}


\section{Introduction}\label{introduction}
The solar radiation is conformed by  electromagnetic waves. It is composed by 
\begin{enumerate}
\item Infrared radiation (IR): It is the cause of the thermal energy, which is classified by
\begin{itemize}
    \item Infrared type A (IRA), in a range of 780-1400~nm.
    \item Infrared type B (IRB), in a range of 1400-3000~nm.
    \item Infrared type C (IRC), in a range of 1~mm to 3000~nm.
\end{itemize} 
\item Visible radiation (Vi): It is the part of the electromagnetic spectrum through the human eye sees. It is in a range of 400-780~nm.
\item Ultraviolet radiation (UV): It is the most energetic part of the radiation. Similar to IR, the UV is classified by:
    \begin{itemize}
   \item UV type A (UVA), in a range of 315-400~nm.
    \item UV type B (UVB), in a range of 280-315~nm.
    \item UV type C (UVC), in a range of 100-280~nm.
    \end{itemize}
\end{enumerate}
The UVC rays are the most energetic, which makes them the most dangerous. However, they are detained in the ozone layer, while the UVB and UVA rays are not fully braked~\cite{WHO}. It is well known that the solar radiation is one of the factors for life on the Earth. It also helps agriculture and other benefits such as on the skin, which helps make vitamin D. However, too much   exposure can cause skin damage~\cite{damage1,damage2,damage3,damage4,damage5}, due to the UV component and the IR causes breakdown of collagen fibers in the skin. However, recently studies have shown that the right combination of IR and IUV can improve skin disorders~\cite{skincare}.\\
It is defined a dimensionless  value for the UV, which provides a prevention of sun exposure, it is called the ultraviolet index (IUV)~\cite{Fioletov2010TheUI, WHO}. It normally is represented in a range from 1 to 11+, where the 11+ means that the upper limit may be greater than eleven, as it is shown in this work. The World Health Organization classifies the values 1 and 2 as low; 3, 4 and 5 as moderate; 6 and 7 as high; 8, 9 and 10 very high and 11 and more values as extreme~\cite{WHO}. For IUV$\leq2$ one can go freely in the outdoors. For IUV values from 3 to 7, it is recommended to use sunscreen, hat and long sleeve shirt, as well as, seek shade at midday. Finally, for the case when the IUV$\geq8$ it is recommended to avoid being outside at midday and it is essential to use sunscreen, hat and long sleeve shirt.\\
As it was mentioned, the ozone layer is a protection to keep out the UVC, over the years it has decreased due to pollution, aerosol, etc. To deplete the ozone layer, the Mont Real protocol was created~\cite{MontReal}. Several studies have been done and they are shown that the UV index has not been change~\cite{UV1,UV2}. Simulations have been done to study the UV index and the effective dose for the production of vitamin D in the human skin~\cite{C5PP00093A}, where it had shown an improvement due to the recovery of ozone layer. These simulations are based on angle radiation  over the Earth~\cite{ozone}.\\
Another relevant characteristic of the Sun (for purpose of this work) is about its magnetic field, which change its direction. This is called Solar cycle. It is related to the sunspots, which decreases and increases around  eleven years~\cite{sunspots}. The effects on Earth due to the Solar cycle are reflected on change climate, radio frequency communications, navigation, among others. Studies  carried out have shown that the Solar cycle occurs between ten and eleven years~\cite{ZHU20233521, unknown, articleChe, SUN2016121, SolarCyclePrediction1, SolarCyclePrediction2}. It has been noted that the maximum of the Solar cycle have occurred at 1993, 2003 and 2013, it is expected that the next maximum value of the cycle occurs in 2025~\cite{2025}. These measurements have been a challenge, because solar radiation on Earth varies due to pollution, urban development and aerosols~\cite{popullation1,popullation2,popullation3,population4,popullation5}.\\\\
In this work, the analysis of data from certain weather stations of Mexico City is carried out, they will simply be referred as stations.  The data acquisition is  explained in Section~\ref{DataAdcquisition}. In Section~\ref{AnalysisResults} is shown the data analysis of UVB index (IUVB) for the months and years from 2000 to 2022. Finally, in Section~\ref{Discussions} the conclusions and discussions are shown.

\section{Data acquisition}\label{DataAdcquisition}
The government of Mexico City has a data base, which contain data of the IUVB values, for each hour since 1998 according to their page ~\cite{SEDEMA}\footnote{At this date, currently the page does not seem to work or is it under repair. The actual page is \url{https://datos.cdmx.gob.mx/ne/dataset/radiacion-solar-uva}, however, when  trying to obtain the data from the year 2,000, it sends to the same page.}. There are 44 monitoring stations, located in various parts of CDMX and the so-called metropolitan area. The information provided is obtained from twelve stations, they are:

\begin{itemize}
    \item Montencillo (MON).
    \item San Agust\'in (SAG).
    \item Pedregal (PED).
    \item Tlalnepantla (TLA).
    \item Merced (MER).
    \item FES Actal\'an (FAC).
    \item Cuajimalpa (CUA).
    \item Santa Fe (SFE).
    \item Laboratorio de An\'alisis Ambiental (LAA).
    \item Milpa Alta (MPA)
    \item Chalco (CHO).
    \item Cuautitl\'an (CUT). 
\end{itemize}
For each year, the data information are collected in ``.xls" files by columns. The first column represents the date of the year, in the second column is the time of day. The following columns are value of the IUV measurements by the stations. Not all stations measured in each year. The data used in this work was from the stations that measured from the year 2000 to 2022, these are MON, SAG, PED, TLA and MER.


The data reported are of two types for ultraviolet solar radiation: 
\begin{itemize}
    \item Solar energy (mW/m$^2$)
    \item Measurement over time (MED/h)
\end{itemize}
 According to~\cite{SEDEMA}, from the measurement over time value, it is possible to obtain the IUVB value, simply multiply the measured value by 2.332~h/MED.
 
\section{Analysis and results}\label{AnalysisResults}
\subsection{Maximum IUVB value distribution} 
\begin{figure}[htbp]
\begin{center}
\includegraphics[width=0.7\textwidth]{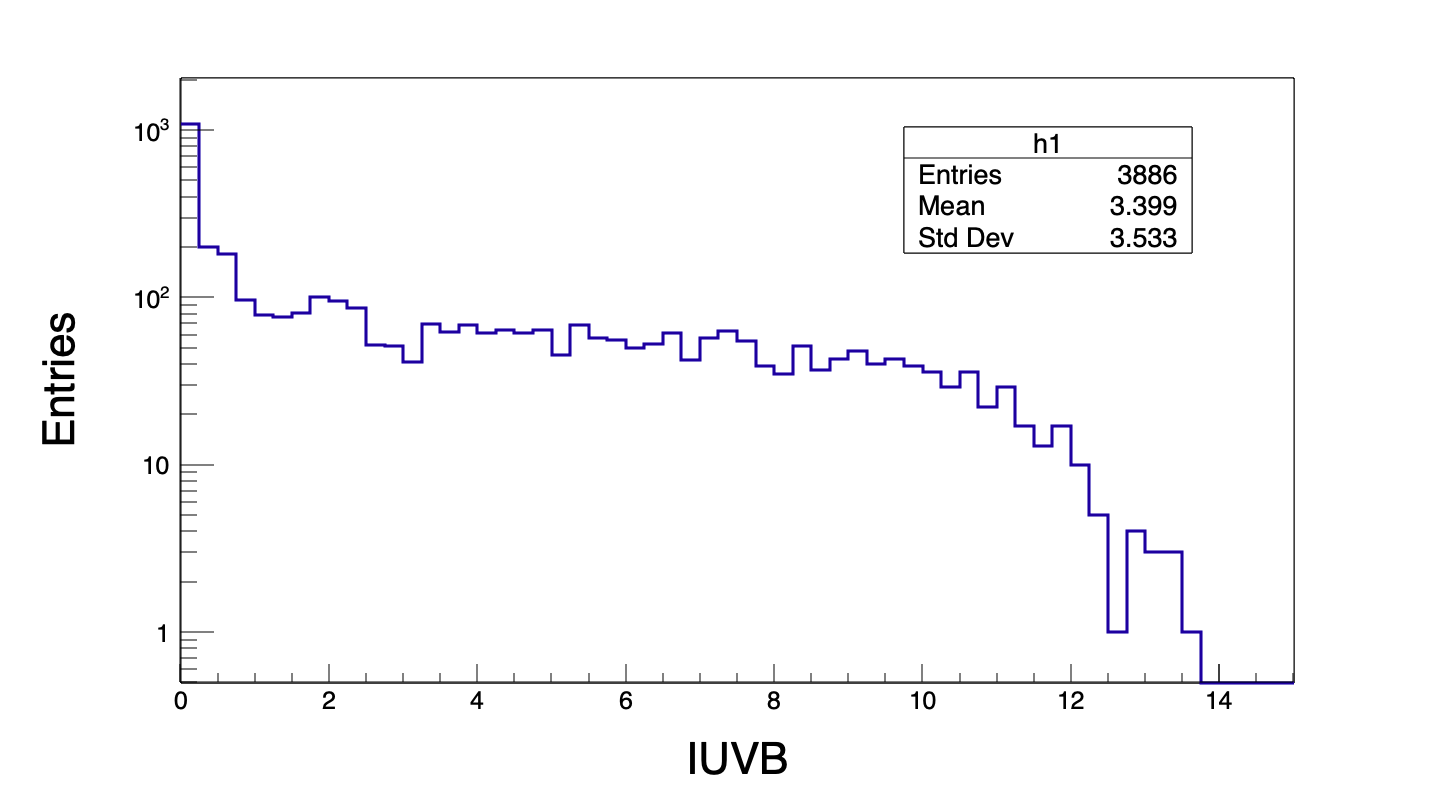}
\end{center}
\caption{IUVB value distribution for the year 2000 for MER station.}
\label{distributionyear}
\end{figure}

\begin{figure}[htbp]
\begin{center}
\includegraphics[width=0.7\textwidth]{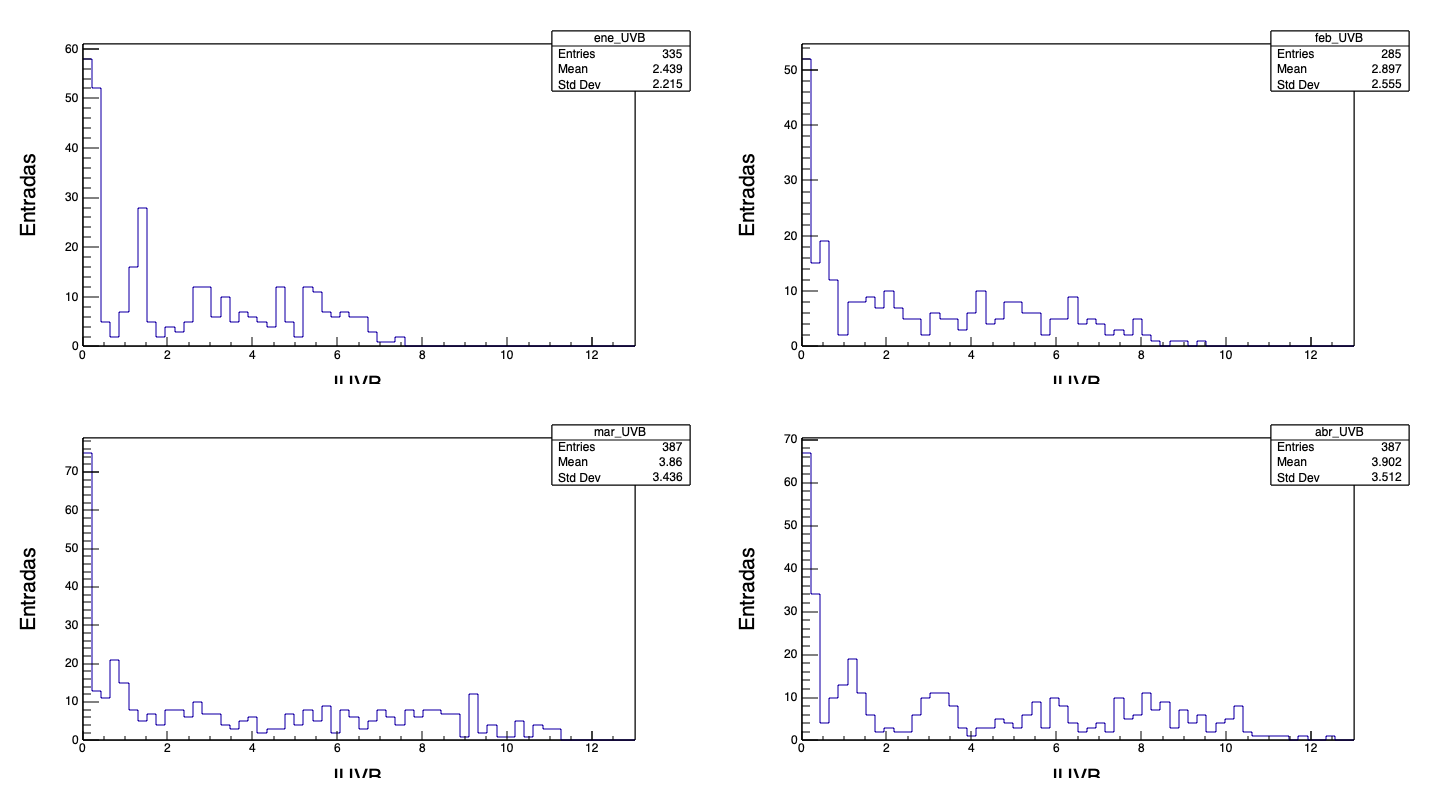}
\includegraphics[width=0.7\textwidth]{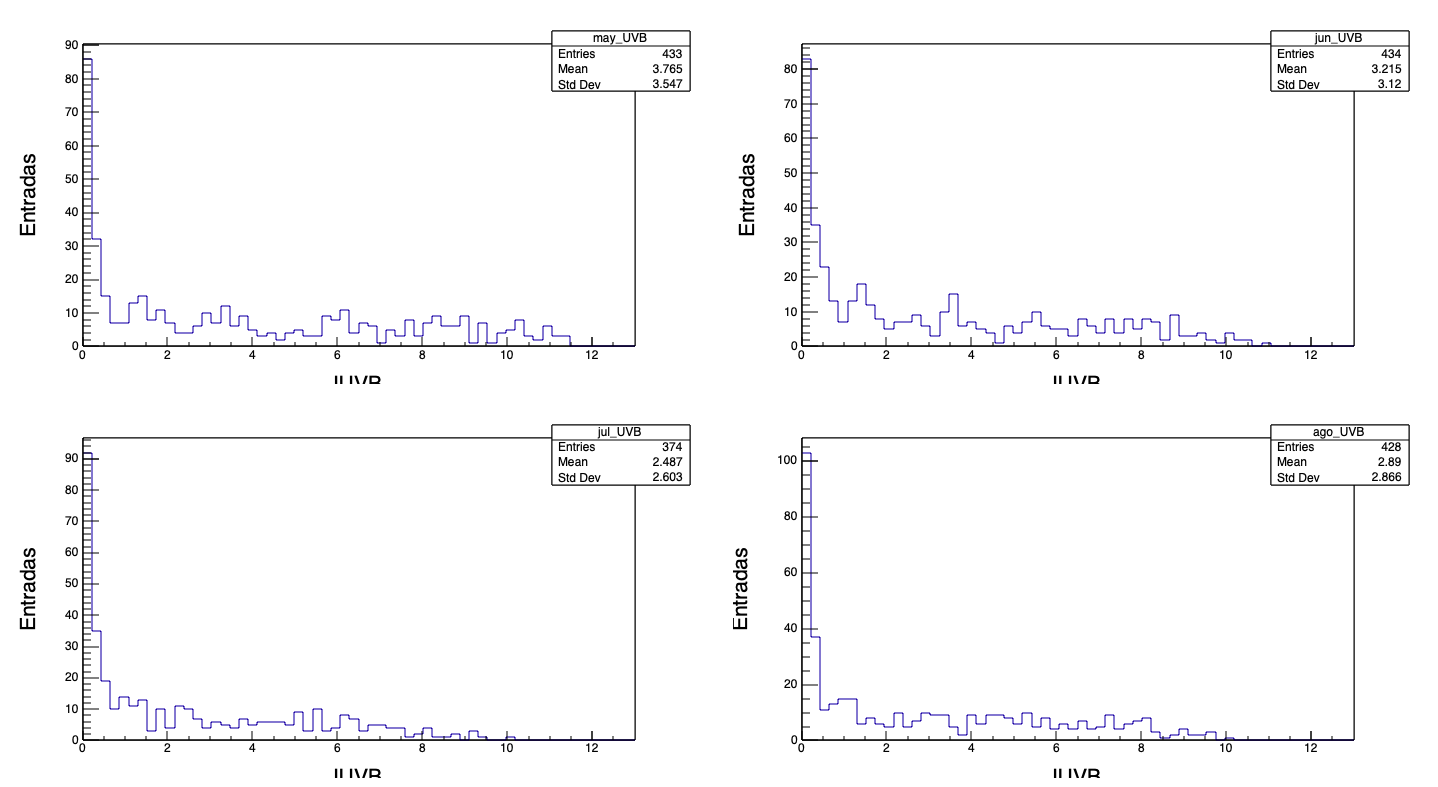}
\includegraphics[width=0.7\textwidth]{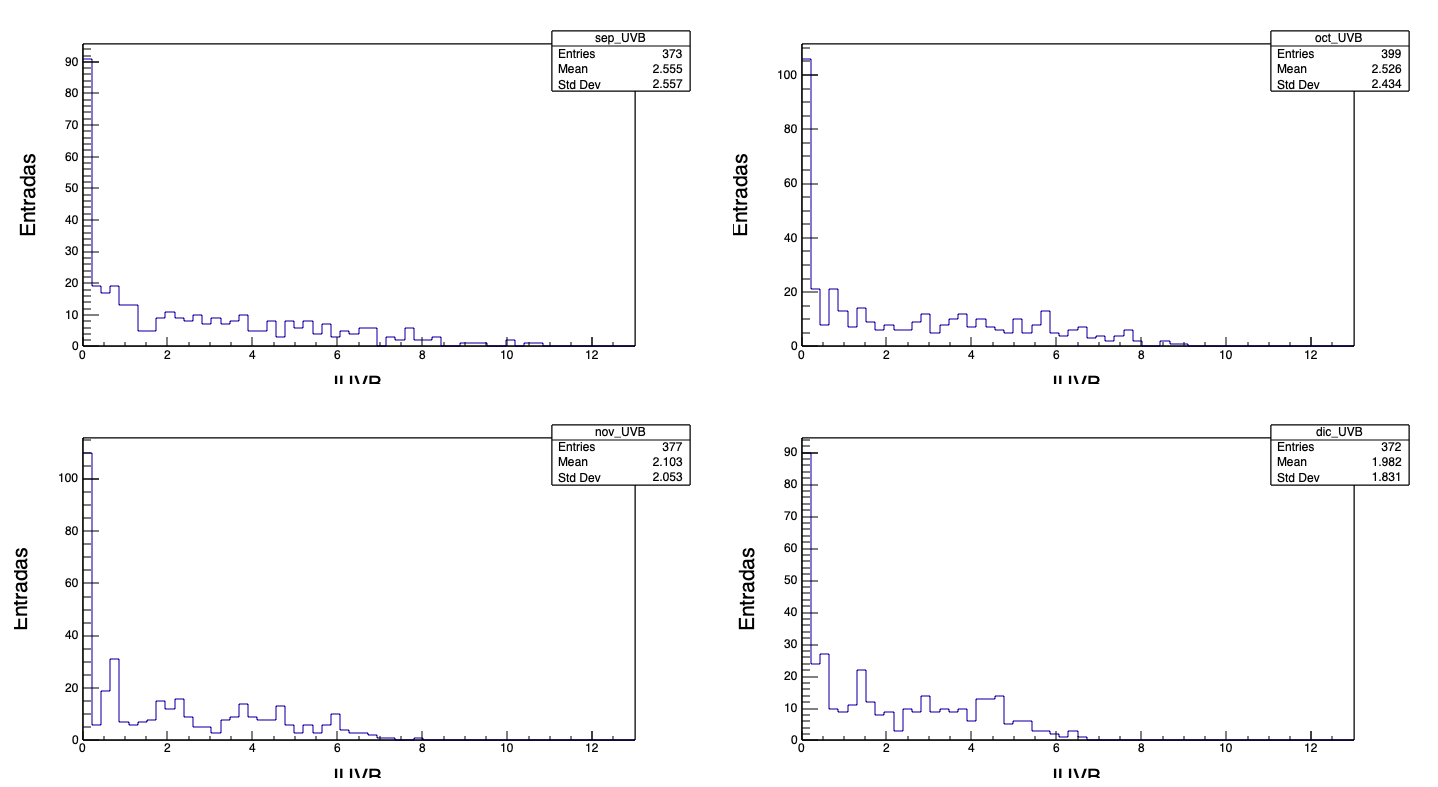}
\end{center}
\caption{IUVB value distribution for the year 2,002. There are shown the twelve months for the MER station.}
\label{distribution}
\end{figure}

The IUVB values distribution per year was obtained, as an example, in Figure~\ref{distributionyear} it is shown the distribution for year 2002 and MER station. From this data, a distribution per month can be obtained, as an example, in Figure~\ref{distribution} it is shown the IUVB value distribution for the MER station for the year 2002. It can be seen that the distribution exceeds the IUV value of nine, for the months March-June. Similar trend was observed for the other stations and for each year.

\begin{figure}[htbp]
\begin{center}
\includegraphics[width=0.7\textwidth]{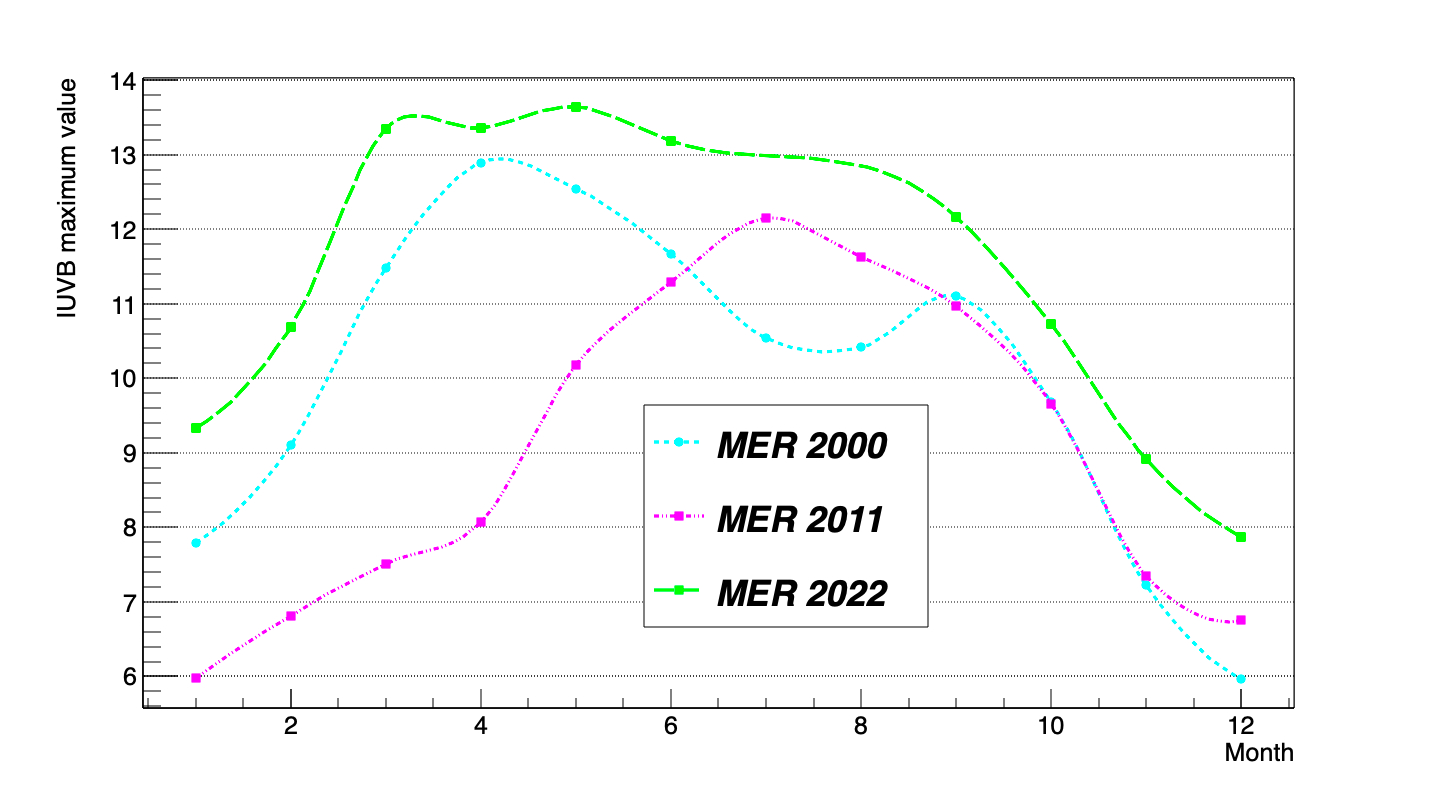}
\end{center}
\caption{Three years maximum IUVB values as function of the month for MER station.}
\label{MaxYear}
\end{figure}

From these distributions it was possible to obtain the maximum value per month. In Figure~\ref{MaxYear}, as an example, it is shown the maximum values distribution for three different years. For the year 2022, it was observed that the IUVB maximum value, in general, it is greater compared with the other years. This type of behavior occurs in the other stations. 

\subsection{Correlation of IUVB value with the time of the day}
\begin{figure}[htbp]
\begin{center}
\includegraphics[width=0.7\textwidth]{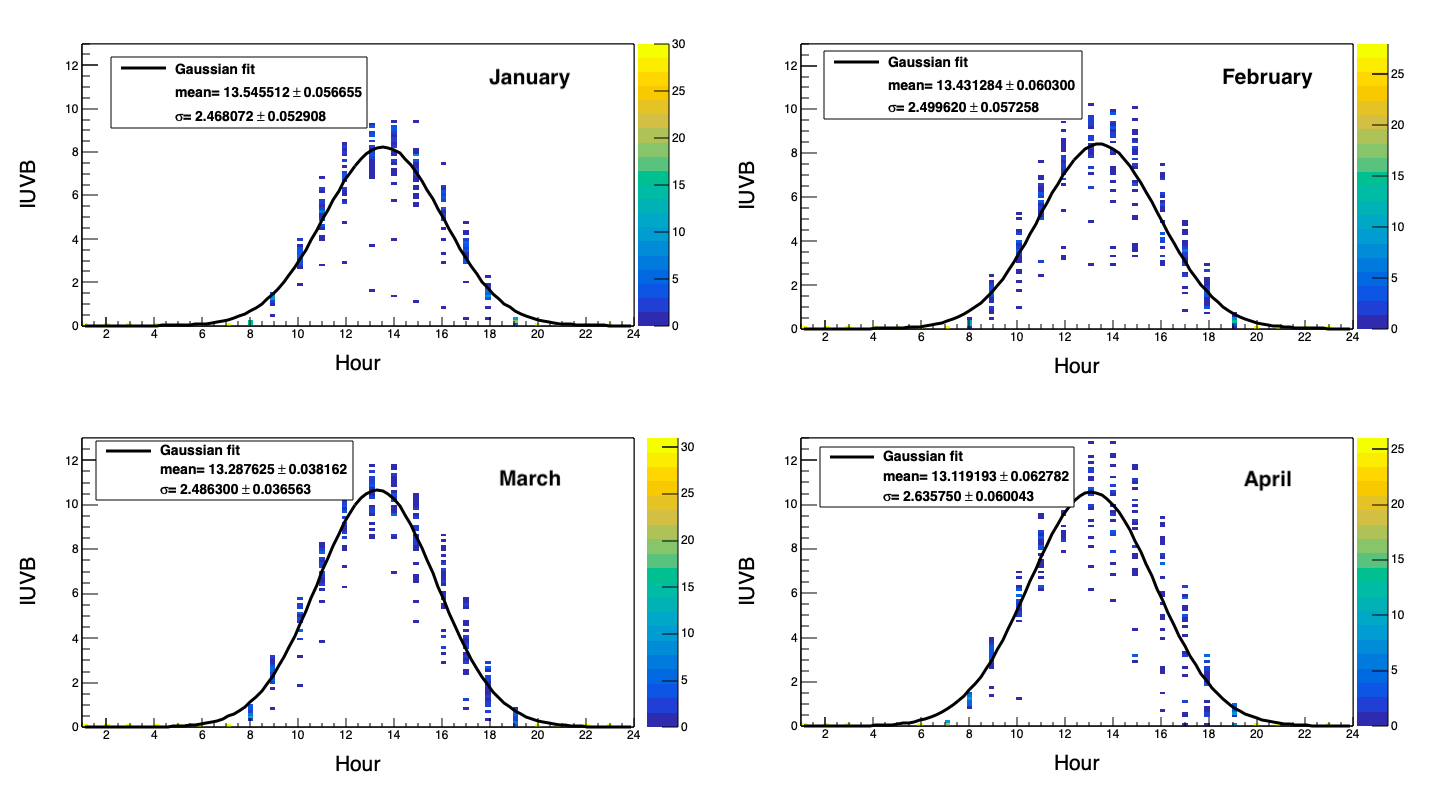}
\includegraphics[width=0.7\textwidth]{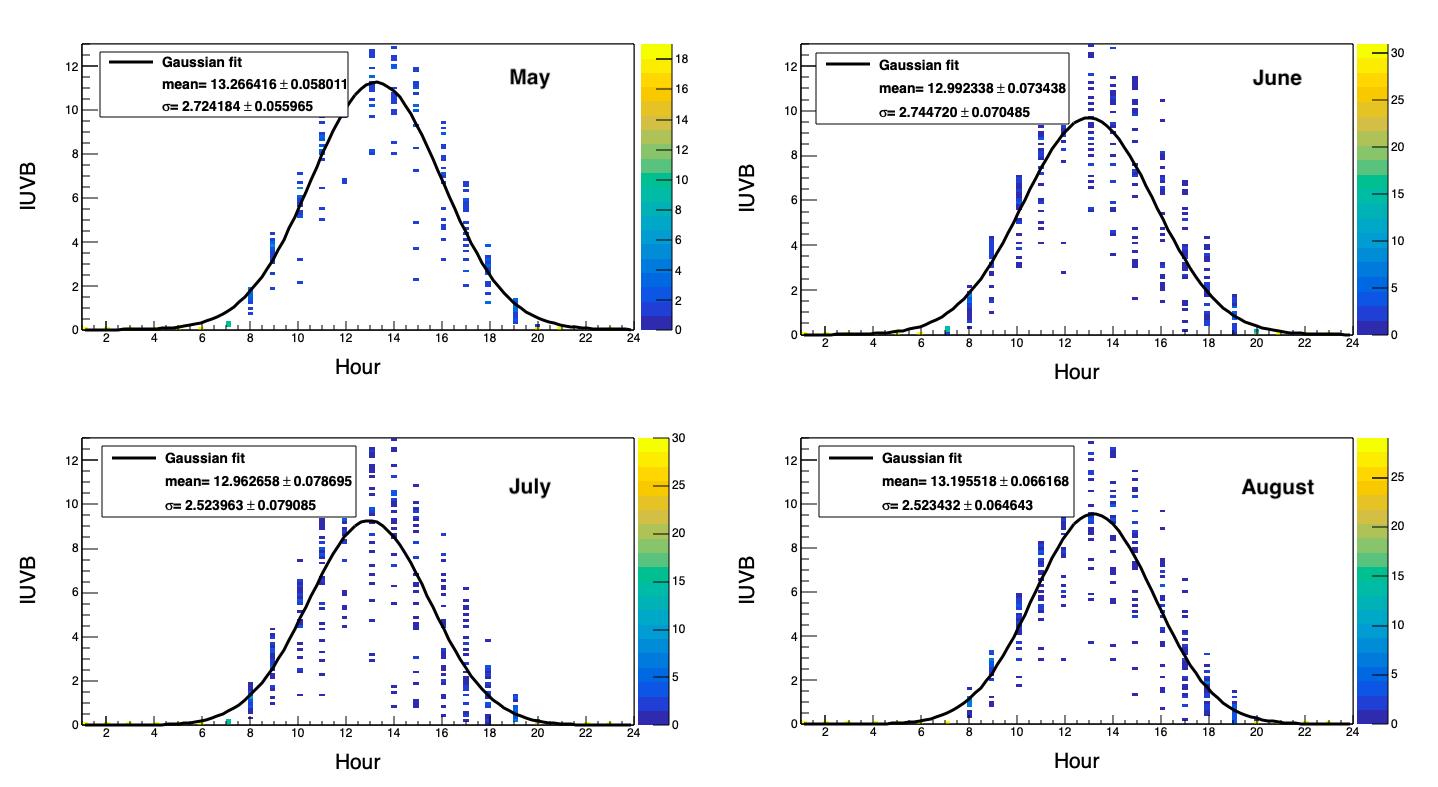}
\includegraphics[width=0.7\textwidth]{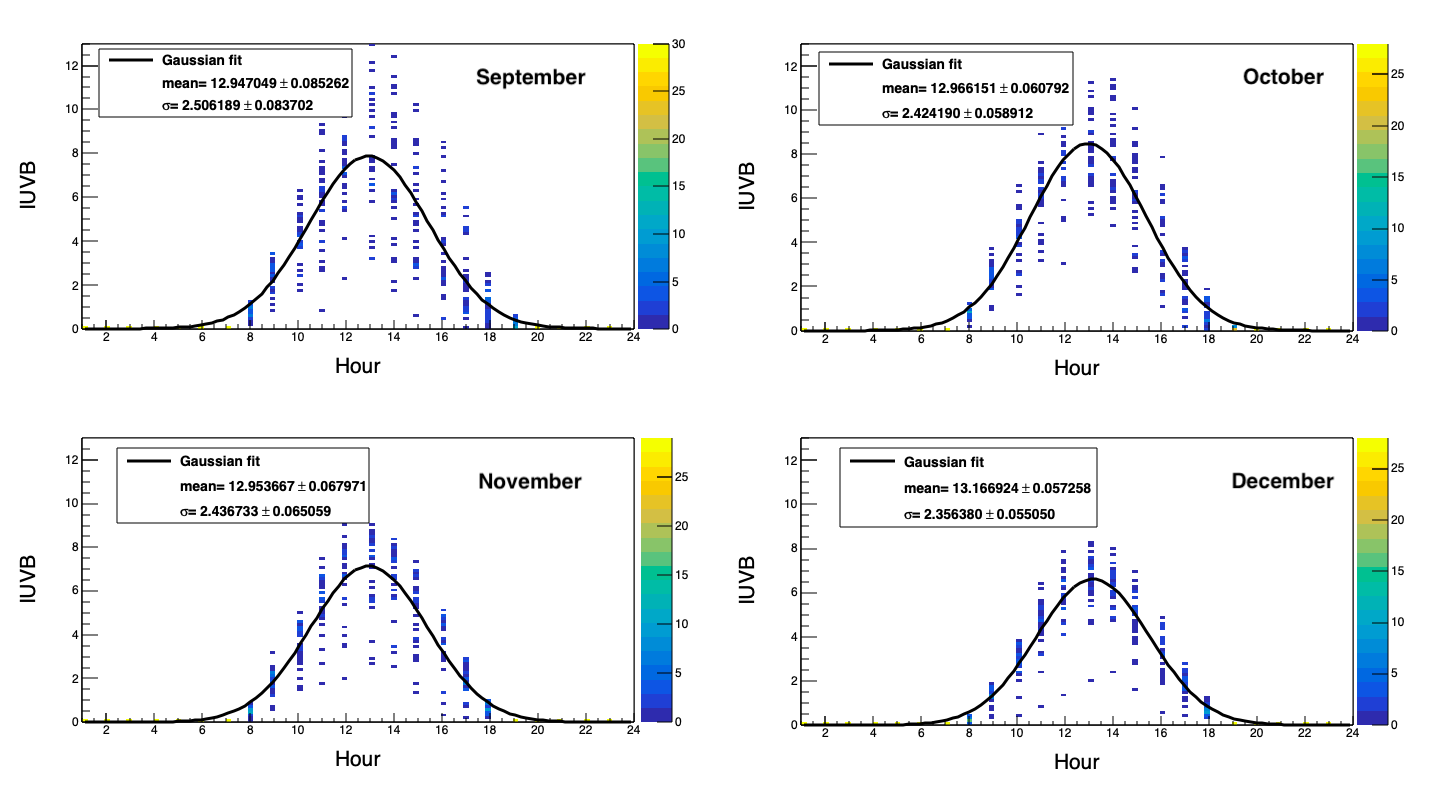}
\end{center}
\caption{IUVB value distribution for the year 2,002. There are shown the twelve months for the MER station.}
\label{IUVvsHour}
\end{figure}

It was possible to obtain a correlation between IUVB value and the hour of day for each month. To illustrate this correlation, in Figure~\ref{IUVvsHour}, it is shown for the twelve months of the year 2002. It can be seen that the number of entries  greater than seven has a tendency to rise and then fall, during the course of the year. This distribution has a Gaussian trend. This behavior occurs in all stations for each year. In general, for the months January and December, the IUV value does not exceed seven, however, between the months February and October, the IUV value exceeds this value. It was observed that the IUVB value exceeds the value of eight for the months of March and August. For the other years and stations, it was observed this same behavior. From the Gaussian fit of these distributions, it was possible to obtain the mean value of all years. The average of these mean values is 13:09~hrs.~$\pm$4~min. This result means that the maximum IUVB value is reached in a range from 13:05 to 13:13.

\begin{figure}[htbp]
\begin{center}
\includegraphics[width=0.7\textwidth]{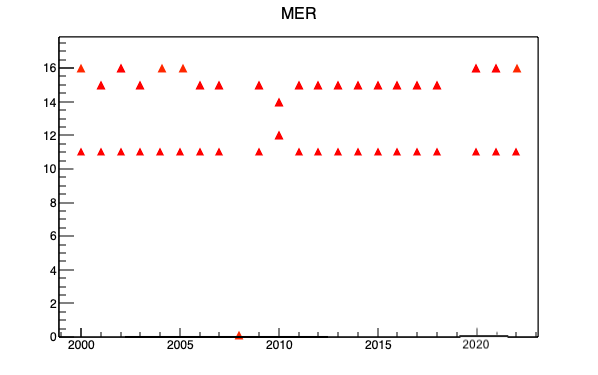}
\includegraphics[width=0.7\textwidth]{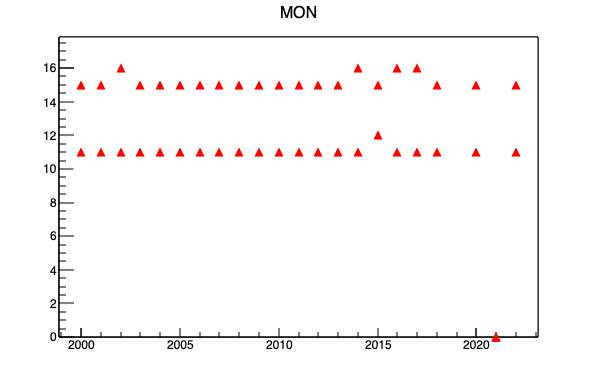}
\end{center}
\caption{Graph made to know the evolution of the interval of the hours of the day over the years, in which the IUV value exceeds seven. It is shown the results for stations MER and MON.}
\label{IntervalTime}
\end{figure}

 Another analysis was performed per month, it consisted of obtaining the range in the hours of the day over the years, for which the UVB values exceed seven. Two of these dependencies are shown Figure~\ref{IntervalTime}, for the MON and MER stations. Based on similar results for the other stations, it was obtained that the interval for which the IUVB values exceed seven is from 11:00 to 16:00.

\begin{figure}[htbp]
\begin{center}
\includegraphics[width=0.7\textwidth]{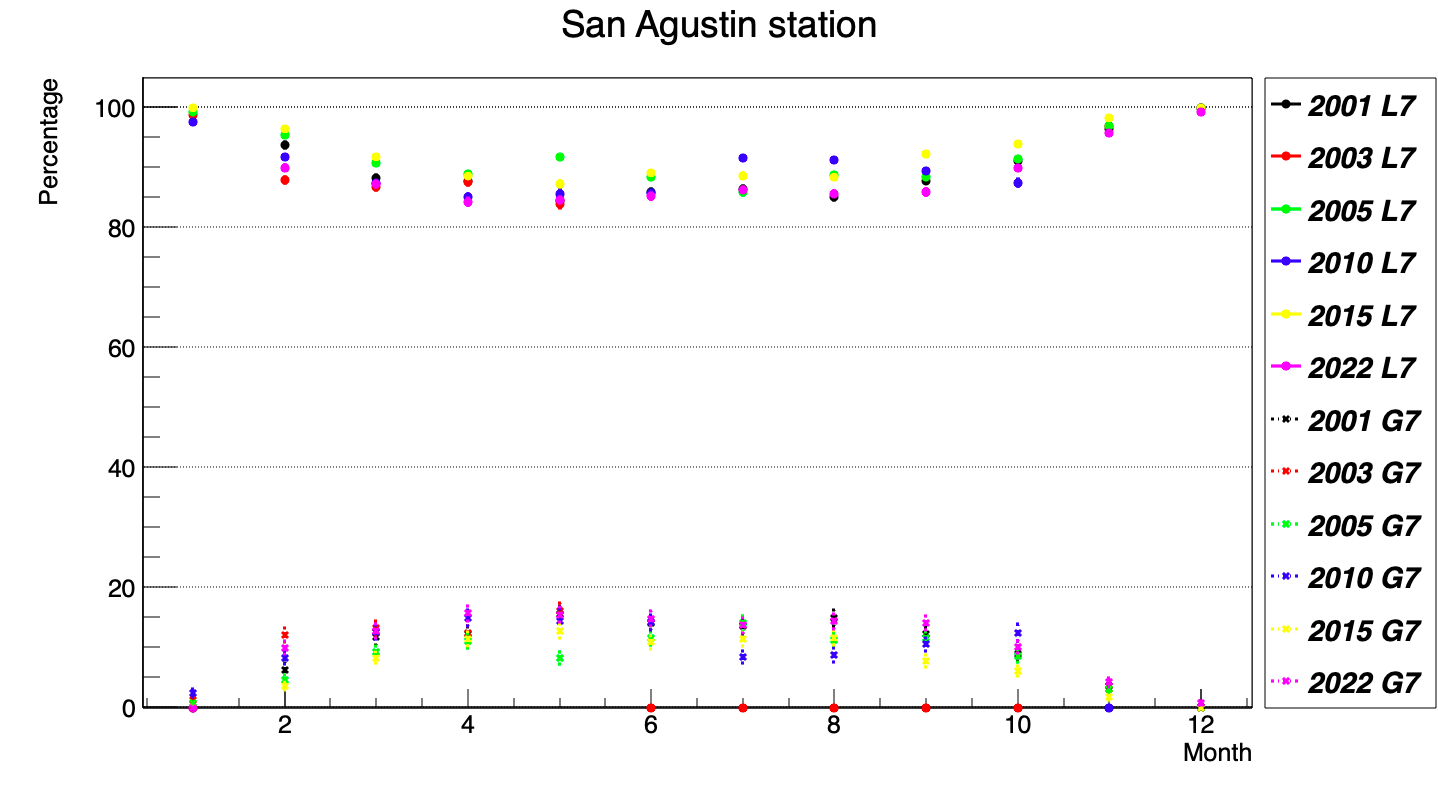}
\end{center}
\caption{Percentage distribution for IUVB values greater (G7 and cross symbol) and less (L7 and circle symbol) than seven. }
\label{G7L7}
\end{figure}

From Figure~\ref{IUVvsHour}, it was possible to obtain the  percentage distribution of the IUVB values greater than seven and the IUVB values  less than seven, as function of the month. As an example, in Figure~\ref{G7L7}, it is shown the UVB distribution for both cases, for the years 2001, 2010 and 2022 for the SAG station.  It can be noted that percentage of IUVB values greater than seven (G7) are less than the 20\%, while the percentage of IUVB values less than seven (L7) are greater than the 80\%. Finally, it can be noted that as G7 increases, L7 decreases. These results are obtained for all years and stations. This trend could have been expected from Figure~\ref{distribution}.

\subsection{IUVB average value}
\begin{figure}[htbp]
\begin{center}
\includegraphics[width=0.8\textwidth]{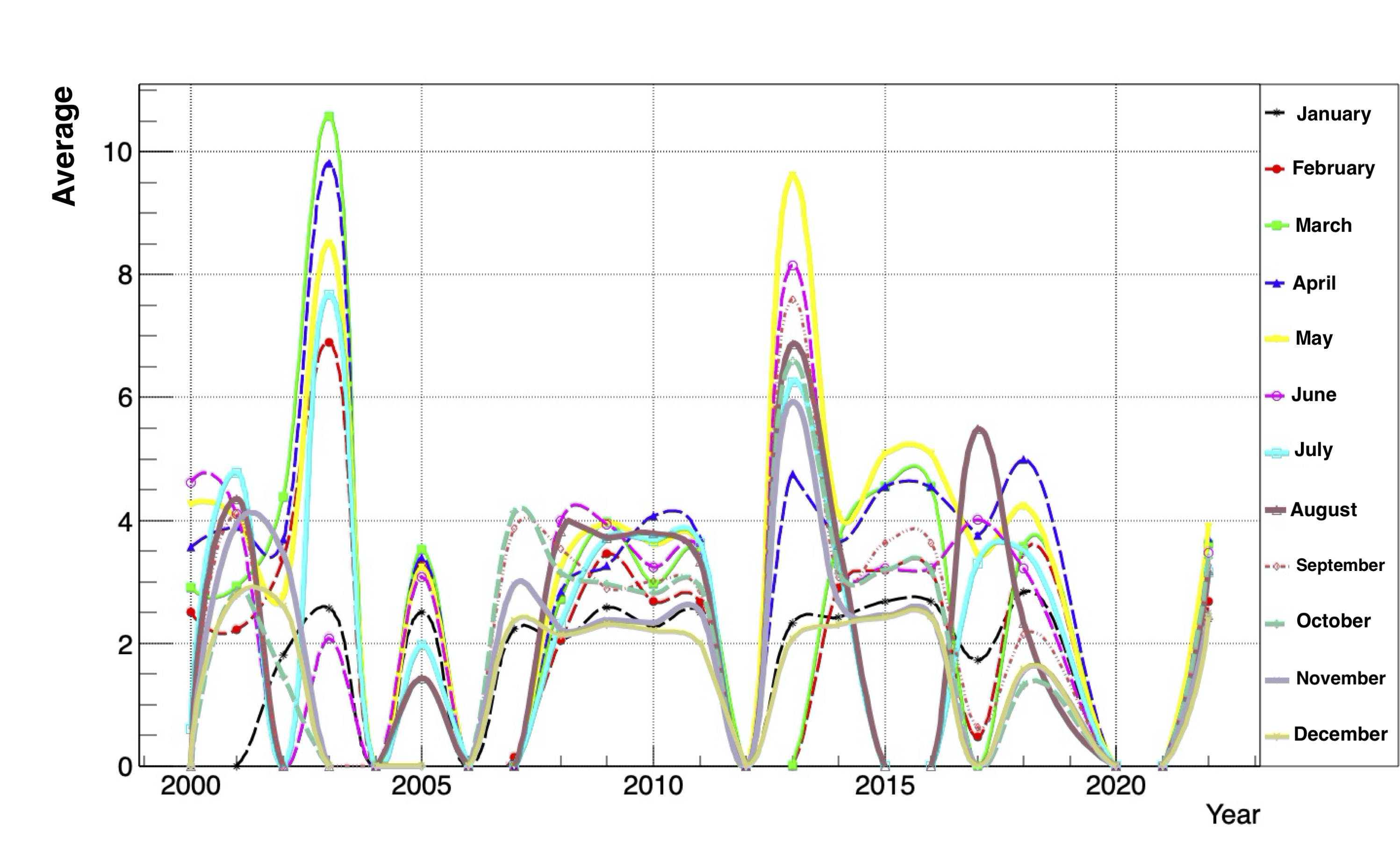}
\end{center}
\caption{Average distribution over the years for the MER station. The years associated with the two maximum peaks coincide when solar cycles have occurred. The zero values means there was no measurement in those years.}
\label{average}
\end{figure}

Obtaining the IUVB average value of each distribution per year, two values are obtained that stand out above the others. To illustrate this result, in Figure~\ref{average} it is shown the average distribution values for the MER station, the two peaks are located at years 2003 and 2013. This trend occurs in a similar way for the other stations. According to solar cycles studies~\cite{ZHU20233521, unknown, articleChe, SUN2016121, SolarCyclePrediction1, SolarCyclePrediction2}, two of them have occurred at 2003 and 2013. Then, the solar cycles were measured by the stations in CDMX through the IUVB values.

\section{Discussion and conclusions}~\label{Discussions}

In this work, an analysis of IUVB solar radiation measured by stations in various points of Mexico City was carried out, of which only five stations were considered, the reason is  because these stations contain information from the year 2000 to the year 2022. Even though data were not collected for certain days and/or months, however, these are the stations with more recorded data.\\
The IUVB distributions per year showed to have fluctuations in data values, as it is shown in Figure~\ref{distributionyear} and more specifically per month as it is shown in Figure~\ref{distribution}. These fluctuations are due to the day and night, because clearly at night the stations does not receive radiation. From these distributions, it was obtained the maximum IUVB value per month, from which it was obtained that distributions of this maximum values fluctuate per year, as it was showed in Figure~\ref{MaxYear} and Figure~\ref{average}. This may be due to different atmospheric factors that affect the measurement, such as pollution~\cite{popullation1,popullation2,popullation3,population4,popullation5}.\\
The IUVB has different values during the daylight hours, at it was shown in Figure~\ref{IUVvsHour}. It was obtained a range in the hours of the day in which the IUVB value is greater than seven (which corresponds to values that are hazardous to health), of the stations tacking into account, it was found that it occurs from 11:00 to 16:00, an example was shown in Figure~\ref{IntervalTime}. From Figure~\ref{IUVvsHour}, can be obtained a range, to consider the mean value and one sigma, which is the 68.2\% of the values, the average range is from 10:38 to 15:41, this interval was obtained to considered all years. From the fit Gaussian distribution, it was obtained the the average mean value is 13:09~hrs.~$\pm$~4~min, which means that in this hour the IUVB reach its maximum value.\\
As it is well known, the solar radiation is dangerous for IUVB values greater than seven. A study was made where it was found that the percentage of the total IUVB values is greater than the 80\%, for the case when the IUVB value is less than seven. While for the case when the IUVB value is greater than seven, the percentage of the total radiation is less than  20\%, as it was shown in Figure~\ref{G7L7}. This result can be intuited from Figure~\ref{distribution}.\\
Finally, it was obtained the IUVB average value for each year. From this average distributions, it was obtained two protruding peaks which are in the 2003 and 2013, as it was shown the Figure~\ref{average}. This two peaks coincide with what reported about when solar cycles have their maximum value~\cite{ZHU20233521, unknown, articleChe, SUN2016121, SolarCyclePrediction1, SolarCyclePrediction2}. The final values distribution in Figure~\ref{average} have an upward trend, similar to the solar cycles studies, mentioned before, which is expected to be in 2024~\cite{ZHU20233521}. In general, the IUVB maximum value has increased, as it can be appreciated in Figure~\ref{MaxYear}.\\
Acknowledgments The authors would like to thank the Secretar\'ia del Medio Ambiente de la Ciudad de M\'exico (SEDEMA), for providing the data in a public manner, in order to perform this work.

\printbibliography

\end{document}